\documentclass[manuscript,screen]{acmart}
\usepackage{tabularx}
\usepackage{cleveref}
\usepackage{booktabs}
\usepackage{longtable}
\usepackage{array}    
\usepackage{hyperref} 
\usepackage{xurl}     

\AtBeginDocument{%
  }

\setcopyright{acmlicensed}
\copyrightyear{2018}
\acmYear{2018}
\acmDOI{XXXXXXX.XXXXXXX}
\acmConference[]{\textbf{Pre-print} Special Issue on Advances in
Social Media Technologies and Analysis ACM Trans. Web}{June 03--05,
  2025}{}
\acmISBN{978-1-4503-XXXX-X/2025/06}

\usepackage{ifthen}
\usepackage{calc}

\newcommand{\annote}[3]{{\color{#3}%
		\colorbox{#3}{\bfseries\sffamily\tiny\textcolor{white}{#2}}
		\color{#3}
		$\blacktriangleright$\textit{#1}$\blacktriangleleft$}
}

\newcommand{\commentFT}[1]{\annote{#1}{FT}{magenta}}

\renewcommand{\annote}[3]{}
\begin{document}

\title{Ethical Risk Analysis of L2 Rollups}

\author{Georgy Ishmaev}
\authornote{Main author. The remaining authors are listed in alphabetical order.}
\email{georgy.ishmaev@inria.fr}
\orcid{0009-0006-6861-1205}
\affiliation{%
  \institution{Univ Rennes, Inria, CNRS, IRISA}
  \city{Rennes}
  \country{France}
}

\author{Emmanuelle Anceaume}
\email{emmanuelle.anceaume@irisa.fr}
\orcid{0000-0003-4158-149X}
\affiliation{%
  \institution{Univ Rennes, Inria, CNRS, IRISA}
  \city{Rennes}
  \country{France}
}

\author{Davide Frey}
\email{davide.frey@inria.fr}
\orcid{0000-0002-6730-5744}
\affiliation{%
  \institution{Univ Rennes, Inria, CNRS, IRISA}
  \city{Rennes}
  \country{France}
}

\author{François Taïani}
\email{francois.taiani@irisa.fr}
\orcid{0000-0002-9692-5678}
\affiliation{%
  \institution{Univ Rennes, Inria, CNRS, IRISA}
  \city{Rennes}
  \country{France}
}



\renewcommand{\shortauthors}{Ishmaev et al.}

\begin{abstract}
Layer 2 rollups improve throughput and fees, but can reintroduce risk through operator discretion and information asymmetry. We ask which operator and governance designs produce ethically problematic user risk. We adapt Ethical Risk Analysis to rollup architectures, build a role-based taxonomy of decision authority and exposure, and pair the framework with two empirical signals, a cross sectional snapshot of 129 projects from L2BEAT and a hand curated incident set covering 2022 to 2025. We analyze mechanisms that affect risks to users’ funds, including upgrade timing and exit windows, proposer liveness and whitelisting, forced inclusion usability, and data availability choices. We find that ethical hazards rooted in L2 components control arrangements are widespread: instant upgrades without exit windows appear in about 86 percent of projects, and proposer controls that can freeze withdrawals in about 50 percent. Reported incidents concentrate in sequencer liveness and inclusion, consistent with these dependencies. We translate these findings into ethically grounded suggestions on mitigation strategies including technical components and governance mechanisms.
\end{abstract}



\received{20 February 2007}
\received[revised]{12 March 2009}
\received[accepted]{5 June 2009}

\maketitle

\section{Introduction}

The blockchain ecosystem has long been praised for its ability to promote decentralization, transparency, and financial autonomy. However, it has also become a fertile ground for ethical concerns, particularly regarding the systematic deception and exploitation of users due to severe information asymmetries. That is to say, many users engage with blockchain applications, without fully grasping the underlying risks and trade-offs. 

From misleading tokenomics and opaque governance structures to security vulnerabilities concealed by project teams, the industry has witnessed numerous instances where users make financial decisions without fully understanding the underlying risks~\cite{mingyi2024}. A good illustration of this issue can be seen in the complex landscape of blockchain Layer 2 (L2) scaling solutions. Many (L2) protocols differ significantly in their design, security guarantees, and governance models~\cite{thibault_blockchain_2022}~\cite{gorzny_rollup_2024}. Users may assume that all L2 solutions offer similar levels of decentralization and trustlessness, while in reality, some require substantial trust in centralized operators. This lack of clarity leads to uninformed decision-making and, in many cases, financial loss~\cite{mingyi2024}.

At the core of this problem lies information asymmetry, a well-documented economic phenomenon in which one party in a transaction has access to significantly more or better information than another~\cite{evans_managers_2017}. Empirical studies show that developers deploying contracts on L2 rollups face a cost versus trust trade off that is difficult to quantify in practice~\cite{lazăr2025ideploycontractspractical}.


The current state of L2 rollups illustrates these issues clearly. At the same time, rollups are a useful case to study how information asymmetries arise, how they create risk, and how they might be mitigated. The slogan ''L2s inherit the security of the underlying blockchain''~\footnote{\url{https://ethereum.org/en/layer-2/learn/}} is a simplification. L2 rollups inherit some immutability guarantees from L1 through state commitments, but only under specific conditions. In practice, L2 security guarantees are conditional and nuanced. We focus on Ethereum L2 rollups as the most mature and diverse ecosystem, and unless stated otherwise ''L2'' refers to Ethereum based systems.

Secondly, high variability and composabilty of L2 components is a good illustration of how seemingly minor architectural variations can lead to dramatically different security and trust assumptions.~\footnote{A trust assumption is a decision regarding the degree to which the supplied indicative (objectively true) properties of domains composing the system should be trusted, and an assessment of the risks associated with wrong evaluation~\cite{haley2004}.} Finally, the problem of information asymmetries received enough recognition (at least in the Ethereum ecosystem) to spur industry initiatives to systematize taxonomies of risks associated with L2s~\cite{Donno2023IntroducingStages}.

A systematic ethical analysis helps us move beyond intuition. Some examples in the introduction are clearly troubling, but many cases are in a ''grey zone'' where intent and responsibility are hard to judge. Today, ethical judgments are mostly made after the fact, not in advance. An ethical risk analysis offers a framework that can be applied proactively to identify problematic design choices~\cite{floridi_ethical_2020}. Its aim is not to assign blame, but to clarify trade offs and highlight better practices in the design of complex systems.

There is still a lack of systematic approaches that link security risks, trust assumptions, and ethical implications arising from information asymmetries in L2 ecosystems~\cite{gorzny_rollup_2024}. Existing dashboards catalog technical risks such as centralized sequencing or upgrade keys. Recent formal security frameworks for L2s classify protocol level assumptions~\cite{Torbagell2025}~\cite{avarikioti2025securityframeworkgeneralblockchain}, but they say little about how those assumptions allocate decision power and exposure across roles. We treat these frameworks as a descriptive baseline, then extend them with an ethical risk analysis that examines how role distributions and trust assumptions create concrete risks for different parties when those assumptions fail.


\begin{quote}
    \textbf{Which operator and governance designs produce ethically problematic user risk in L2 rollups?} 
\end{quote}	

We adapt ethical Risk Analysis (eRA) framework to L2 rollups, build a role-based taxonomy of risk allocation, and pair it with two empirical signals, a cross sectional snapshot of 129 projects from L2BEAT and a hand curated incidents dataset from 2022 to 2025. We find ethically problematic architectural and governance choices affecting users' withdrawals in around half of projects. We use these results to identify problematic role distributions in L2 systems using eRA methodology and propose mitigation priorities for operators and developers.
\\
\\
Section \ref{sec:tech_back} provides a technical background and explains key architectural components of L2 solutions. We explain the methodology of analysis and data sources in section \ref{sec:method}. We explain the theoretical framework and motivation for its choice in Section \ref{sec:theor_framework}, as well as its general application in the context of this research. Section \ref{sec:ethical_analysis} provides a detailed ethical analysis of L2 solutions in the context of eRA application, and external validation on the basis of aggregated data for the prevalence of potential risk in different L2s, as well as historical track data on relevant L2 incidents. We discuss the limitations of the analysis and provide recommendations in Section \ref{sec:discussion}. 

\section{Technical Background}
\label{sec:tech_back}

Blockchains such as Ethereum~\cite{buterin2013ethereum} support asset transfers and smart contracts\footnote{A smart contract is simply a program that runs on the Ethereum blockchain. It is a collection of code (which provides the smart contract’s functionality) and data (which correspond to the smart contract’s state) that reside at a specific address on the Ethereum blockchain.}, but only at a few dozen transactions per second~\cite{bez2019}. Security and decentralization choices cause high fees and congestion during peak activity. To increase capacity without weakening core guarantees, several scaling techniques have been proposed, including state channels, Plasma, and rollups, collectively termed Layer 2 (L2)~\cite{thibault_blockchain_2022}. Built on top of Layer 1 (L1), they offload transaction processing while relying on L1 for ordering and data availability. L2s differ from sidechains, which are separate blockchains with their own validators and consensus protocols\cite{song2024}.

Currently, rollup based L2s dominate by user and market share~\cite{gorzny_rollup_2024}, so we focus on their components and functionality.

\begin{figure}[h]
\centering
\includegraphics[width=0.7\linewidth]{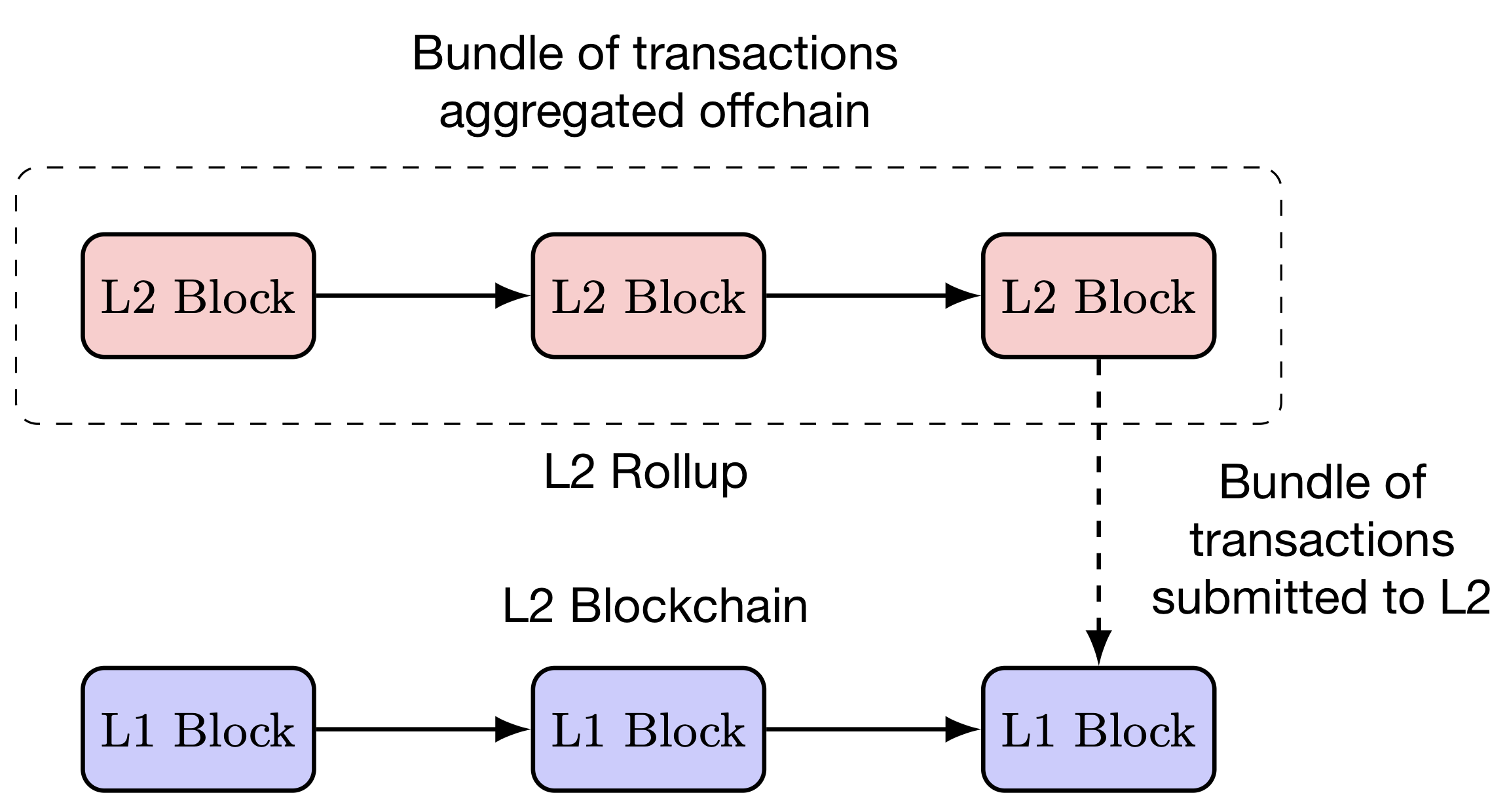}
\caption{Data compression of a rollup. L2 rollups collect multiple transactions off-chain and compress them into a compact bundle (or ‘‘roll up’’). This consolidated data, often accompanied by a cryptographic proof, is then periodically submitted to the L1 blockchain, allowing it to finalize and secure the rollup’s transactions under the base layer’s consensus. Finalization is the point in Ethereum at which a block is considered economically irreversible by consensus, so rational validators will not reorganize it and any L2 commitments included in it can be treated as permanent for exits and verification.}
\label{fig:rollup_data}
\Description{}
\end{figure}

Rollups bundle many transactions and publish the compressed data on Ethereum (Figure~\ref{fig:rollup_data}). They maintain an independent state tracked by a dedicated smart contract. By committing minimal data to Ethereum, they reduce L1 computation.

In a rollup, a specialized operator, a single node or coordinated components, executes transactions and updates state~\cite{palakkal_sok_2024}. The operator aggregates user transactions, compresses them, and submits a rollup block to Ethereum. ZK rollups attach a validity proof, while optimistic rollups assume correctness unless challenged via fraud proofs. We use operator as an umbrella term for entities across rollup designs and governance, as detailed in subsections 2.1. and 2.2.

There are also emerging proposals for ''native'' or ''enshrined'' rollups, where some or all of the verification logic is moved from smart contracts into the L1 protocol itself. From a user perspective these designs aim to narrow the gap between L1 and L2 security by reducing reliance on off chain operators~\cite{gorzny2026early}. A detailed analysis of native rollups is beyond the scope of this paper, as we focus on currently predominant rollup solutions.

We next decompose a generic rollup. A rollup is not a fixed architecture or protocol, it is a method to connect an L2 ledger to L1, enabling verifiable commitment of L2 transactions to L1.

\begin{figure}[ht]
\centering
\includegraphics[width=0.9\linewidth]{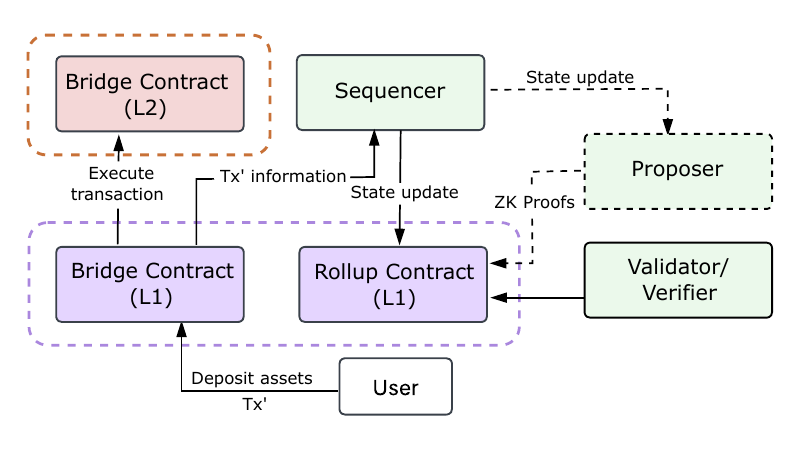}
\caption{Generic rollup architecture. A user’s transaction \textit{Tx'} (e.g., a deposit) is collected by a sequencer, bundled with other transactions, and committed to the L1 as an updated state. In a ZK rollup (dashed arrow), the proposer collects state update from a sequencer and submits a cryptographic proof attesting to state correctness. Whereas in an optimistic rollup, the proposed state can be challenged by validators through a fraud-proof process.}
\Description{Roles and role overlaps in eRA framework.}
\label{fig:rollup_arch}
\end{figure}

\subsection{Smart Contracts}
\label{sec:smart_contracts}

Verifiable commitments are implemented with smart contracts on EVM compatible blockchains~\cite{zheng_vm_2020}. Modularity and composability lead to diverse implementations~\cite{thibault_blockchain_2022}~\cite{palakkal_sok_2024}, so this architecture reflects current practice. Smart contracts enable L2s to leverage L1 security.

Most rollups split functionality across contracts, as in Figure~\ref{fig:rollup_arch}. One or more contracts handle rollup state, tracking L2 state roots, verifying proofs, or managing fraud challenges, while separate contracts handle deposits, withdrawals, and message bridging between L1 and L2\footnote{At a high level, data transfer between L1 and L2 resembles inter-contract communication on Ethereum, with some distinctions. It is facilitated by a pair of specialized “messenger” smart contracts, one on each respective layer, which abstract low-level communication details, similar to how HTTP libraries abstract network protocols.}. We refer to these as the \emph{Rollup contract} and \emph{Bridge contract}.

\textbf{Rollup contract} is the L1 smart contract that maintains rollup state, enforces validity rules, and facilitates deposits and withdrawals. It stores a Merkle root~\cite{merkle1987digital} committing to the current state, which includes balances, code, and other L2 data.

How state commitments are verified defines two rollup types. The key problem is ensuring the posted state root is correct, otherwise a submitter could alter state and move assets.

\begin{itemize}
\item \textit{Optimistic Rollups} assume batches are valid, but permit fraud proofs during a challenge window, typically on the order of a week on Ethereum mainnet, to dispute invalid updates~\cite{sun_doubleup_2024}.
\item \textit{ZK Rollups} use cryptographic proofs verified on L1, so there is no challenge period. A \textit{prover} generates a proof that the new state follows from correctly applying the transition function to the sequenced block, the contract verifies it and updates the state root~\cite{chaliasos_analyzing_2024}.
\end{itemize}
Terminology varies. In optimistic systems, \textit{validators} challenge fraudulent batches. In ZK systems, the L1 contract verifies proofs, so no separate validator set is needed. Calling the \textit{prover} a validator is misleading, since it produces proofs for the L1 contract.\footnote{These also should not be confused with validating nodes in Proof of Stake (PoS) consensus protocols, elsewhere in the text referred to as~\textit{PoS validators} for the sake of clarity } The roles of validators and provers are discussed in Section~\ref{sec:rollup_operators}.


\textbf{Bridge contract} is the cross chain communication component, also called the canonical bridge. L1 native assets are locked in the bridge on origin, then equivalent tokens are minted on L2, creating canonical representations within the rollup. Withdrawals move assets back from L2 to L1 securely.



\begin{figure}[h]
\centering
\includegraphics[width=0.8\linewidth]{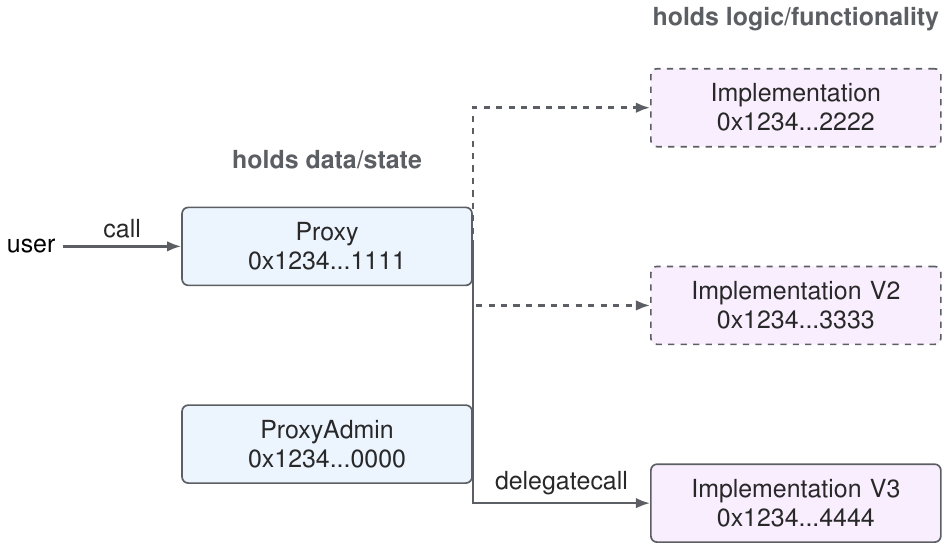}
\caption{Proxy pattern for smart contracts. The proxy contract collects user calls and delegates them to a separate ‘‘implementation’’ contract that holds the actual logic. This design enables contract upgrades without changing the address users interact with.}
\Description{}
\label{fig:proxy}
\end{figure}

In practice, rollups use many contracts, not just two. For example, the Optimism (OP) rollup6 has 16 smart contracts on Ethereum mainnet\footnote{\url{https://docs.optimism.io/stack/rollup/overview}}, including multiple bridges, message passing contracts, configuration registries, and several dispute and execution contracts.
Proxy contracts add more. A proxy delegates calls to an \textit{implementation contract}~\cite{iii_proxy_2023} Figure~\ref{fig:proxy}. Users call the proxy, which forwards requests while keeping a stable address. This enables upgrades and rollbacks, but also introduces deployment complexity and the risk of malicious upgrades~\cite{iii_proxy_2023}.

This decomposition reflects common components in deployed rollups. Early designs used a single contract for fraud proofs and deposits, creating a single point of failure. Modern designs separate rollup and bridge logic across multiple L1 contracts with corresponding L2 contracts. As we show next, however, functional decentralization alone does not prevent \textit{de-facto} centralization.

When we refer to L1 security we mean that, even if L2 operators fail or censor, users can reconstruct the L2 state from data posted on Ethereum, verify it, and exit to L1 without trusting any committee. The strength of this guarantee depends on the rollup’s data availability (DA)~\footnote{For a rollup to be secure, all data needed to reconstruct the L2 state must be available.} choice, for example using blob transactions (''blobs'') on Ethereum or an external DA layer.

\subsection{Rollup Operators}
\label{sec:rollup_operators}

It is often said that rollups inherit the security and trust guarantees of the Layer 1 they are built on, and they do in a general sense. In practice, these guarantees vary across implementations. An L2 chiefly inherits L1 security by (1) publishing the rollup’s state root or validity proofs to L1, and (2) holding user assets in L1 contracts that reference those roots. Anchoring the state root in the L1’s tamper resistant ledger makes L1 consensus protect against unauthorized state changes.

These guarantees depend on the specific composition of rollup components, as outlined in Section \ref{sec:smart_contracts}, and on the roles of rollup operators. By \textit{operator} we mean a single node entity or an entity comprising multiple components, responsible for functions within rollup architectures. Common roles are: \textbf{Sequencer}, \textbf{Validator} (optimistic rollups), \textbf{Prover} (ZK rollups), and \textbf{Verifier}.

In practice most rollups are monolithic, where a single trusted party (often the project team handles sequencing, bridging, and verification of state on the L1 chain~\cite{gorzny_rollup_2024}. In principle, these roles can be modularized and controlled by distinct operators~\cite{motepalli_sok_2023}, but our analysis in Section \ref{sec:ethical_analysis} shows that functionally decomposed modules are often operated by the same entities. 

\textbf{Sequencer} orders and bundles transactions, sometimes executes them, and submits batches to L1. Its core tasks are ordering and aggregation~\cite{derka_sequencer_2024}. Implementations range from decentralized sets to (for emerging shared-sequencer designs see~\cite{bearer2024espresso}~\cite{han2025_a_layer}) a single entity, yet most current rollups use a centralized sequencer~\cite{derka_sequencer_2024}. For simplicity, we treat the sequencer as a single functional node that may encompass services like \textit{batcher} and \textit{proposer}, which respectively publish L2 data batches to L1 and publish new state roots from executed blocks. A \textit{centralized sequencer} means it is maintained by the rollup team, runs on centralized servers, and uses one account to post batches to L1.

A centralized sequencer gives near instant confirmations to L2 users and reliably submits valid batches to L1, but introduces two key risks. First, a failure halts L2 transaction processing and batch submission~\cite{li2025_denial_attacks}~\cite{chaliasos2025unaligned}. Second, as sole transaction router, it controls ordering and can extract maximum value from reordering (MEV) {\footnote{MEV refers to the maximum amount of value a blockchain miner or POS validator can make by including, excluding, or changing the order of transactions. For example, to front run transactions to decentralized exchanges.}}. The user risks and guarantees can diverge from L1. A trusted sequencer may block external MEV bots common on L1, yet it can internalize these costs~\cite{alipanahloo2024}~\cite{torres2024}. MEV on L2s is less explored than on L1, with recent work on cross-rollup MEV and fast-finality effects~\cite{gogol_cross-rollup_2024}. One difference is opacity, since centralized trusted sequencing limits access to L1 style mitigation tools for L2 users.

In some rollups, instead of a single sequencer, \textbf{Validators} can propose new rollup blocks, similar to Ethereum nodes. To participate they stake funds in the L1 rollup contract, and selection probability is proportional to stake. Validators can submit state data to the rollup contract and challenge others. Malicious behavior, such as submitting invalid state data, leads to slashing. However, many current optimistic rollups rely on fraud proofs from predetermined whitelisted entities rather than a permissionless validator set formalized in the protocol.

In ZK rollups, \textbf{Provers} may be centralized under the sequencer’s operator or independent~\cite{chaliasos_analyzing_2024}. Security wise, one entity operating sequencer and prover still cannot arbitrarily alter funds or state, as long as L1 verification and DA guarantees hold and cannot be bypassed by governance, since invalid updates are rejected by the L1 contract~\cite{Chaliasos_2025}. However, a centralized operator can censor transactions, and if a centralized sequencer or proposer fails, withdrawals from L2 to L1 may be blocked. A permissionless prover set can add resilience against a misbehaving or faulty sequencer~\cite{stephan2025crowdprove}. Only a fully permissionless set of sequencers and provers can approach L1 equivalent security, liveness, and trust~\cite{motepalli_sok_2023}, though doing so requires additional consensus, and given ZK proving costs it remains open whether such designs preserve throughput and cost advantages.
\textbf{Verifier} is the L1 onchain contract or component that checks the correctness of rollup state commitments and governs when they are finalized. In \textit{optimistic} rollups it adjudicates fraud proofs during the challenge window and can slash invalid claims, in \textbf{ZK rollups} it verifies submitted validity proofs before accepting new state roots, so this role exists in both models.

In practice, operators are mostly companies that develop and run L2s, sometimes with non profit entities as foundations controlling upgrade keys, emergency powers, and who is allowed to operate what. Individuals sometimes do participate, but mainly as security council signers \footnote{A Security Council is a small threshold multisig~\footnote{Short for multiple signatures wallet, a blockchain wallet that requires approval from multiple private keys.} committee, appointed by a foundation or company, that can execute critical or emergency actions such as contract upgrades}, independent validators or challengers, and, increasingly, as external provers, but they are not the typical entity running the production sequencer.
\begin{enumerate}
  \item \textbf{Core project companies} typically run production \textit{sequencers} and often \textit{provers} today, for example OP Stack chains describe a single party sequencer, and Arbitrum documents its sequencer role.\footnote{\url{https://docs.optimism.io/stack/rollup/overview}}\footnote{\url{https://docs.arbitrum.io/how-arbitrum-works/sequencer}} Base is incubated by Coinbase.\footnote{\url{https://docs.base.org/}}
  \item \textbf{Foundations and decentralized autonomous organizations (DAOs)} usually control \textit{upgrade keys} via multisigs or Security Councils, which can appoint or replace operators, for example OP Stack’s security model and the Arbitrum DAO governance docs.\footnote{\url{https://docs.optimism.io/stack/security/faq-sec-model}}\footnote{\url{https://docs.arbitrum.foundation/gentle-intro-dao-governance}}
  \item \textbf{Independent participants} increasingly contribute as \textit{validators/challengers} and external \textit{provers}, for example Arbitrum’s BoLD enables permissionless validation, and zkSync is opening its prover via APIs and third party networks.\footnote{\url{https://docs.arbitrum.io/how-arbitrum-works/bold/gentle-introduction}}\footnote{\url{https://zksync.mirror.xyz/1HzbsDVMQeE3P-DgysjKzagNnHehwCIrC1QA9XcNRnk}}\footnote{\url{https://www.lagrange.dev/blog/a-new-era-for-zk}}
\end{enumerate}

In the following sections we outline the aspects that most impact user risk and compare implementations using an existing dataset of key L2 architectural differences.

\subsection{Operational and Governance Risks}
\label{sec:op_gov_risks}

Governance risks in L2 rollups are a class of hazards that arise from discretionary control by operators over critical components that determine inclusion, finalization, upgrades, and data release. These components include the sequencer, state root proposer or batch poster, prover and verifier, the canonical bridge, and upgrade keys. 

From the perspective of standard security taxonomies, these are mostly additional socio-technical assumptions about how operators behave and how governance works, not assumptions about the underlying cryptography~\cite{avarikioti2025securityframeworkgeneralblockchain}~\cite{Torbagell2025}. We, however, analyze them as first-class, because as later sections show  these account for most of the ethically relevant risks to users.

Appendix \ref{app:incident_glossary} provides the incident classes observed across Layer 2 rollups from data provided in Appendix \ref{app:historic_data}. Every class is applicable to both Optimistic and ZK rollups because both rely on off chain operators for ordering or proving, on L1 contracts for settlement and bridging, and on governance controlled configuration that can constrain or disable user pathways. \emph{Variability}, explains how the same class manifests in Optimistic and ZK rollups. Thus for the purpose of this study we do not analyze different types of rollups specifically.

\section{Methodology}
\label{sec:method}

Our evaluation is combining data analysis with the Ethical Risk Analysis (eRA) framework proposed by Hansson~\cite{hansson_ethical_2017,hansson_how_2018}\commentFT{I'd add `proposed by Hansson~\cite{hansson_ethical_2017,hansson_how_2018}'} to assess the security and ethical dimensions of L2 rollups. We treat existing formal security frameworks for L2s~\cite{avarikioti2025securityframeworkgeneralblockchain}~\cite{Torbagell2025} as a catalog of technical assumptions, and use eRA to analyze how those assumptions, and their failure, distribute risks and benefits across different operator and user roles.

\subsection{Data Collection}

\textbf{Data sources and provenance}. We analyze Layer 2 rollup projects using a snapshot of the public L2BEAT scaling summary API, saved as \texttt{summary\_l2beat.json}. Python code in jupyter notebook together with data snapshot are available online at \url{https://gitlab.inria.fr/gishmaev/l2-ethical-risk-analysis/-/tree/main}

\textbf{Schema exploration and key discovery}. Before extraction, the script enumerates object paths and prints representative samples, which surfaces field presence, optionality, and example values. We include projects whose category is in {ZK Rollup, Optimistic Rollup, Other}. Field names are lowercased and trimmed, categorical values are standardized, and one row per project is produced with a stable \texttt{project\_id key}. For each included project, the script traverses the risks array and extracts five dimensions reported by L2BEAT. We record the project identifier, the risk name, the categorical value, the sentiment label, and the short description. 

\textbf{Historic incidents dataset}. We compile historic incidents from two sources without relying on an incidents API. First, we scrape each project page at \url{https://l2beat.com/scaling/projects/}. Second, we add incidents identified in public reports from official project websites and social media posts. After de duplication and cleaning, the dataset contains 32 incidents spanning June 2022 through August 2025. For each incident we record project, start date in UTC, short description, incident type category, source URL, and whether the source is L2BEAT or an external report. Short version of table is provided in Appendix \ref{app:historic_data} Full historic data table is available at \url{https://gitlab.inria.fr/gishmaev/l2-ethical-risk-analysis/-/tree/main}.


\subsection{Conceptual Analysis Methodology}

We explain motivation and for the choice of specific theoretical framework Ethical Risk analysis (eRA) in the next section. Here we provide step-by-step methodology of eRA, that is applied in the following sections.

\begin{figure}[h]
  \centering
  \includegraphics[width=0.5\linewidth]{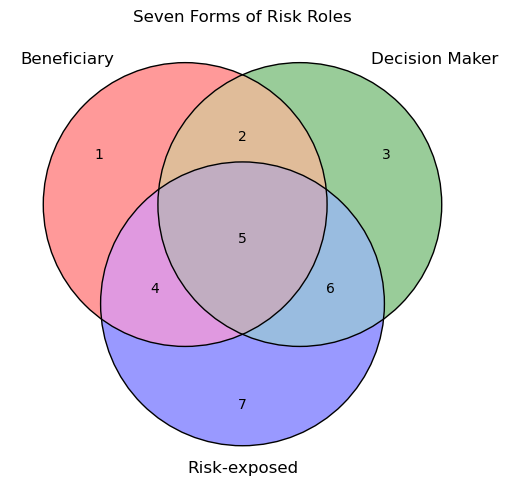}
  \caption{Roles and role overlaps in eRA framework. Adapted from Hansson S.'How to Perform an Ethical Risk Analysis (eRA)'~\cite{hansson_how_2018}.}
  \Description{Roles and role overlaps in eRA framework. Adapted from Hansson S.'How to Perform an Ethical Risk Analysis (eRA)'~\cite{hansson_how_2018}.}
  \label{fig:era}
\end{figure}

On the basis of technical analysis provided in \ref{sec:tech_back}, and using eRA framework [25], we identify the roles through which risk is generated, allocated, and governed, namely risk exposed persons, beneficiaries, and decision makers (Step 1). We focus on risks discussed in \ref{sec:op_gov_risks}. The taxonomy is descriptive, it tests whether those who profit also share burdens and whether decision makers are exposed to the consequences of their choices [17]. We then map role overlaps with a Figure \ref{fig:era} to detect ethically salient configurations, notably Field 2, decision and benefit without commensurate exposure, which encourages risk taking and dilutes accountability, and Field 7, exposure without benefits or decision making (Step 2). We also note Field 1, benefit only, and Field 4, exposure and benefit without decision power. Finally, eRA structures a detailed ethical deliberation, individual risk benefit weighing, distributional analysis, rights analysis, and power analysis, to assess fairness, consent, and governance (Step 3).

\section{Theoretical Frameworks and Risk Analysis}
\label{sec:theor_framework}

This section presents analysis of L2 systems risks through the prism of the eRA framework, namely Step 1 'role analysis' and Step 2 'role distribution'. Step 3 of eRA application - 'detailed ethical deliberation' - is introduced in the separate following section \ref{sec:ethical_analysis}.

The eRA framework is useful for L2 rollups for two reasons. First, it builds on standard security analysis rather than replacing it. Security analysis alone does not fully capture power imbalances, information asymmetries, and incentives that shape how risks are distributed. In particular, it does not explain why it is morally concerning when rollup operators, who control essential components, offload risks onto less informed users. eRA adds a normative lens that makes these trust assumptions and role distributions explicit.

Second, eRA focuses on the roles of different entities and participants, and provides clear criteria for identifying ethically problematic relations and dependencies. This is especially important in decentralized systems where dependencies can be dynamic and emergent. In L2 rollups, many configurations are possible, with operators taking different roles, which in turn determines different trust assumptions (Appendix \ref{app:incident_glossary}).


Security analysis characterizes technical threats, vulnerabilities, and controls, often with probabilistic treatment (see also\cite{avarikioti2025securityframeworkgeneralblockchain} for a general L2 security framework). Its outputs are necessary, but not sufficient, for ethical assessment in socio-technical systems. Security analysis is largely agent agnostic, it rarely asks how risks are distributed or who can unilaterally change them, and it does not assess whether risk allocation is acceptable when control and benefit asymmetries are large. We therefore apply a relational and normative taxonomy of risk distribution to rollup specific risks, as discussed in Section~\ref{sec:op_gov_risks}.


\subsection{Role Identification and Classification (eRA Step 1)} 

As explained in Section \ref{sec:rollup_operators} different L2 configuration can present different dependencies. However, we can generalize some role distribution patterns prevalent in current landscape of L2 architectures. Summary is given in Table \ref{tab:era_roles}.

End users are the primary \textbf{risk exposed} group. They face financial and operational risks that include asset freezing, censorship, theft through invalid state finalization, liquidity constraints that block exit during stress, and exposure to malicious or negligent upgrades. End users are heterogeneous in expertise and use case. A simple case involves a payment on a rollup, while developers who deploy applications are also users whose funds and operations can be affected by settlement or upgrade behavior. Independent validators or watchers who stake funds bear operational risks as a function of their active participation in the architecture. We distinguish these from ordinary users because their operational engagement changes the risk profile.

\textbf{Beneficiaries} include rollup operators and sequencers who receive fee revenue and may extract MEV, governance groups such as multisigs and security councils that benefit through token appreciation, fee structures, and influence, and Rollups as a Service providers that profit by simplifying deployment for operators who lack deep infrastructure expertise. Users, including developers, are also beneficiaries, for example through lower fees or faster settlement, which matters for the Field 4 analysis below.

\textbf{Decision makers} include centralized operators, multisig or security council governance groups, and core L2 developers who control transaction inclusion, system upgrades, and protocol architecture. Decisions by developers who manage protocol design and security patterns carry high impact, even in the absence of formal role labels. RaaS providers act as de facto decision makers by selecting defaults and abstractions that shape downstream risk, for example when a default disables or restricts escape mechanisms. Developers of the underlying L1s influence the feasible design space of L2s and therefore indirectly affect L2 risk allocation. Users, in most configurations, lack direct influence over design and upgrades. Where permissionless governance or DAOs are nominally present, empirical practice frequently concentrates decision making in a small group [14].

\begin{table}[t]
\centering
\caption{Role classification in L2 rollups}
\label{tab:era_roles}
\begin{tabular}{lccc}
\toprule
roles & Risk exposed & Beneficiaries & Decision Makers \\
\midrule
End users & yes & yes & no \\
Application developers as users & yes & yes & no \\
Independent validators or watchers & yes & limited/no & no \\
Rollup operators & indirect & yes & yes \\
Sequencers & indirect & yes & yes \\
Governance groups (multisigs, security councils) & indirect/no & yes & yes \\
Rollups as a Service providers (RaaS) & indirect/no & yes & yes \\
Core developers & indirect/no & indirect/no & yes \\
L1 developers & no & no & indirect \\
Independent provers & indirect/no & yes & no \\
\bottomrule
\end{tabular}
\end{table}

\subsection{Role Combination and Dependencies (eRA Step2)} 

The eRA approach highlights two ethically problematic combinations.

\textbf{Field 2, beneficiaries and decision makers without commensurate risk exposure.} Operators, sequencers, and governance groups often make security relevant decisions, for example around upgrade timing and content, while collecting fees and potential MEV, and while being insulated from the largest downside that falls on users. RaaS providers can fall into this category when their architectural defaults set risk relevant parameters for operators without the providers bearing user facing losses. This configuration incentivizes risk taking that is not aligned with the interests of those who bear the risk, it undermines \textbf{justice and fairness}, and it weakens \textbf{accountability} because the primary cost bearers lack levers over decisions.

\textbf{Field 7, only risk exposed.} Many end users are risk exposed without meaningful influence or adequate protection. Canonical cases include the inability to withdraw funds when sequencers fail and no emergency exit is available, or when an upgrade modifies core contracts immediately and without time locks, making exit practically impossible before new risks materialize. This is an ethically vulnerable configuration, it raises concerns about \textbf{rights}, in particular the right not to have risks imposed without recourse, and it cuts against \textbf{non maleficence}.

Two additional fields warrant structured scrutiny.

\textbf{Field 1, only beneficiary.} Clear instances are rarer in rollups since most gains are tied to either decision power or some operational exposure. However, independent provers that are paid per proof request can approximate Field 1, assuming sound implementation and absent material counterparty or operational risk. The ethical question here concerns whether and to what extent it is acceptable for some parties to benefit from risks borne by others, which implicates \textbf{justice and fairness} even if direct harm is not apparent.

\textbf{Field 4, risk exposed and beneficiary without decision power.} Many users accept lower fees and faster settlement in exchange for elevated risk from upgrades, censorship, or exit constraints, while lacking any direct influence over governance. If the realized benefits meaningfully exceed the risks, the position may be substantively favorable, however procedurally it remains deficient because decisions are taken by others. Edge cases blur Fields 4 and 7, for example when the apparent benefit, a single cheaper transaction, is small while the tail risk from bridging funds to an L2 is large, which effectively places the user closer to Field 7.

We take the last step of eRA structured ethical deliberation aligned to principles, in the next section and combine it with identified risks and additional quantitative analysis.

\section{Ethical Analysis}
\label{sec:ethical_analysis}
\subsection{Core principles from IT and cybersecurity ethics}

We adopt a set of well established principles from professional IT and cybersecurity ethics as the normative touchstone for the analysis that follows~\cite{flechais2023practical}~\cite{ACM2018code}. We omit the topic of privacy as even though it is a very prominent ethical issue in blockchain application it is out of the scope of this paper.

\begin{itemize}
    \item \textbf{Beneficence, public good.} Security activities ought to promote user well being and societal welfare, for example by choosing designs that make digital systems safer for non expert users and by minimizing foreseeable harm to the broader ecosystem.
    \item \textbf{Non maleficence, do no harm.} Designers and operators should avoid causing harm through action or omission. Where foreseeable harms exist, for example because of upgrade mechanisms or liquidity dependencies, they should be mitigated with appropriate controls and clear user facing safeguards.
    \item \textbf{Integrity and honesty.} Claims about system properties and guarantees should be truthful and appropriately qualified. Communications to users should not obscure material limitations, residual risks, or conditions under which stated guarantees may fail.
    \item \textbf{Justice and fairness.} Risks and benefits should be distributed in a manner that does not systematically disadvantage those with the least power or decision making. Evaluation should include whether costs are externalized to parties who cannot influence risk creating decisions.
    \item \textbf{Accountability.} Decision makers should be answerable for security relevant outcomes. Mechanisms such as auditable change processes, time locked upgrades, and ex ante oversight support traceability and responsibility.
    \item \textbf{Professional competence.} Actors should exercise due care and diligence, applying state of the art controls proportionate to risk, and avoiding negligent errors that stem from insufficient preparation or expertise.
\end{itemize}

These principles provide the ethical vocabulary that eRA will operationalize when we analyze concrete role constellations and design choices in L2 rollups.

\subsection{Structured ethical deliberation aligned to principles (Step 3)}

\textbf{Individual risk benefit weighing.} 
Users often cannot make informed risk benefit trade offs because key information is underspecified or inconsistent across rollups. Instant or unannounced upgrades, incomplete or non functioning fraud proof mechanisms, and whitelisted proof systems limit informed choice. Independent resources, such as L2Beat, reduce information asymmetry, but heterogeneous designs and unclear requirement traceability still make comparison difficult. The lack of standardized safety requirements, for example for escape hatches and emergency withdrawal windows, prevents ordinary users from assessing whether a rollup’s trade offs are acceptable~\cite{gorzny_rollup_2024}. These conditions conflict with integrity and honesty, since user facing claims often omit qualifications, and they undermine professional competence when state of the art safety practices are not adopted.

\textbf{Distributional analysis.} 
The distribution of risks and benefits is skewed. Retail users bear much of the serious downside risk, for example sudden fund freezes or theft after malicious upgrades, while centralized operators and governance bodies receive fee revenue, MEV, and other upside~\cite{gogol2025priorityfailsrevertbasedmev}. Shared sequencing also introduces non obvious constraints on arbitrage extraction~\cite{silva2025shortpaperatomicexecution}. Institutional actors can often mitigate risk through preferential access and operational sophistication, for example preemptive withdrawals, while retail users cannot. This asymmetry conflicts with justice and fairness. A full quantitative assessment is beyond scope, but the structural features above make the direction of the imbalance clear.

\textbf{Rights analysis.}  Users voluntarily accept some risks, for example small delays or ordering compromises, in exchange for lower costs. At the same time they are involuntarily exposed to significant risks that arise from governance and upgrade mechanisms. Instant upgrades without time locks, exit windows, or community oversight expose users to new hazards without an opportunity to withdraw, which is analogous to imposing risk without consent. Operating without functioning fraud proofs while advertising fraud based security, or restricting proof systems to whitelisted actors, curtails user autonomy and undermines the spirit of the claimed guarantees. These practices conflict with \textbf{non maleficence} and \textbf{rights} based constraints on risk imposition.

\textbf{Power analysis.} Decision authority in many L2 rollups is concentrated in small groups, for example multisig governance bodies, centralized sequencers, and core development teams. These actors can reorder or censor transactions, upgrade contracts, or change security relevant parameters without meaningful community consent or user recourse. Checks and balances are limited, for example the absence of time locked upgrades or mandatory exit windows, which weakens \textbf{accountability}. Concentrated power, combined with information asymmetry, entrenches the Field 2 and Field 7 configurations identified above (Figure\ref{fig:era}). These arrangements increase risks for end users and call for governance designs that redistribute decision authority, constrain unilateral action, and standardize safety mechanisms.

\subsection{External validation}
\label{sec:external_validation}

We conduct external incident-based validation by analyzing historical failures incidents in L2s and a comparative sample of similar systems to corroborate and refine the risk assessment. We combine two data sources. First, Table \ref{tab:l2b_risks_analyzed} provides descriptive prevalence of operator risk conditions across 129 L2 projects. Second, Table \ref{tab:historic_incidents} records 32 publicly documented incidents between 2022 and 2025 across 22 L2 projects, labeled by type of incident. There is no direct one-to-one mapping between the historic incident labels in Table 3 and the governance driven risk categories in Table \ref{tab:l2b_risks_analyzed}, therefore we treat these datasets as complementary signals. We report descriptive shares only and make no causal claims.


\begin{table*}[t]
  \caption{Architectural Risks from L2Beat Risk Data. Snapshot July 2025, categories as defined by L2Beat.}
  \label{tab:l2b_risks_analyzed}
  \centering
  \setlength{\tabcolsep}{6pt}
  \renewcommand{\arraystretch}{1.1}
  \begin{tabularx}{\textwidth}{@{} l >{\raggedright\arraybackslash}X r r @{}}
    \toprule
    Type & Potential risk & Projects & Share \\
    \midrule
    State validation & No state validation, the system permits invalid state roots. & 32 & 24.8\% \\
    Exit window & There is no window for users to exit in case of an unwanted regular upgrade since contracts are instantly upgradable. & 111 & 86.0\% \\
    Proposer failure & Only the whitelisted proposers can publish state roots on L1, so in the event of failure the withdrawals are frozen. & 65 & 50.4\% \\
    Sequencer failure & There is no mechanism to have transactions be included if the sequencer is down or censoring. & 17 & 13.2\% \\
    Data availability & Proof construction and state derivation rely fully on data that is NOT published onchain. & 35 & 27.1\% \\
    \multicolumn{2}{r}{\textbf{Total projects analyzed}} & \textbf{129} & \textbf{100.0} \\
    \bottomrule
  \end{tabularx}
\end{table*}

\textbf{Descriptive prevalence}. As provided in Table \ref{tab:l2b_risks_analyzed}. Across the 129 projects, the most common governance-driven operator risk is the absence of an exit window due to instant upgrades, present in 86.0 percent of projects. Proposer failure that can freeze withdrawals is present in 50.4 percent. Data availability external to L1 is present in 27.1 percent. State validation that is not enforced at L1 appears in 24.8 percent. Sequencer failure with no credible forced inclusion path is present in 13.2 percent. These prevalences describe potential for harm, not realized incidents.

\textbf{Incident distribution}. Looking at the historical tracking of incidents, Table \ref{tab:historic_incidents}, we can see 32 incidents from June 2022 to August 2025. Incidents concentrate in four coarse failure types. (1) Sequencer liveness or inclusion incidents account for 59.4 percent of records; (2) withdrawal or bridge incidents account for 18.8 percent; and (3) exploit or security bug failures account for 12.5 percent; (4) Censorship or forced inclusion failures 9.3 percent. The single most frequent label is sequencer outage, downtime, with 19 occurrences. 

Consistency and divergence between the two signals. The datasets align qualitatively on the centrality of operator controlled liveness and finalization paths, but they diverge where latent governance hazards have not yet produced recorded incidents.
\begin{itemize}
	\item Sequencer liveness and inclusion. Table \ref{tab:historic_incidents} shows frequent sequencer related incidents. Table \ref{tab:l2b_risks_analyzed} reports that 13.2 percent of projects lack a credible forced inclusion path. The apparent tension is resolved once we distinguish between presence of a nominal forced inclusion mechanism and its practical usability, (L2BEAT data flag is a nominal indicator but not sufficient proxy for forced inclusion usability).  Many incidents involve outages or congestion where a forced path either existed but had parameters or operational requirements that made it impractical for ordinary users and relayers, or where finality still depended on other centralized components. Ethically, this places users in Field 7 configurations (Figure \ref{fig:era}), risk exposed without decision making capacity, despite formal mitigations. The remedy is to evaluate forced inclusion designs for operational readiness rather than presence, including documented parameters, public relayers, and tested fallbacks.
	\item Proposer liveness and withdrawal finality. Table \ref{tab:l2b_risks_analyzed} indicates that 50.4 percent of projects are vulnerable to proposer failure that can freeze withdrawals. Table \ref{tab:historic_incidents} includes withdrawal and bridge incidents, 18.8 percent of incidents, that illustrate the user-facing consequences of liveness dependencies in the finalization pipeline. While labels differ, the ethical pattern is consistent with Field 2 and Field 7 interactions (Figure \ref{fig:era}), where small groups hold decision power over finality while users bear the cost of freezes.
	\item Exit windows and upgrade discretion. The 86.0 percent prevalence of instant upgrades without exit windows in Table \ref{tab:l2b_risks_analyzed} is not mirrored by a large number of recorded incidents in Table \ref{tab:historic_incidents}. This is expected, because absence of time locks is a latent hazard that becomes salient under adversarial or negligent upgrades or under emergency responses that change contracts. The lack of recorded incidents does not lower the ethical salience, given the magnitude of harm such events would impose on users who cannot withdraw under the old rules. eRA therefore assigns high priority to the mitigation of this risk category irrespective of current incident counts.
	\item State validation and data availability. The prevalence of optional or bypassable state validation at L1 and the use of external data availability, 24.8 and 27.1 percent respectively, are not strongly reflected in Table 2 because their failures are harder to observe and are often shielded by operator controlled processes. Invalid state finalization and data withholding produce catastrophic risks that may be rare or undisclosed. eRA treats these as high severity risks with structural implications for accountability and rights, even when incident data are sparse. 
\end{itemize}

\begin{table}[t]
  \caption{Historical distribution of L2 incidents by type for 2022-2025}
  \label{tab:historic_incidents}
  \centering
  \setlength{\tabcolsep}{6pt}
  {%
\small
\begin{tabular}{lrr}
\toprule
Incident type (compressed) & Count & Share (\%) \\
\midrule
Sequencer disruptions (halt, outage, performance) & 19 & 59.4 \\
Bridge or withdrawal issues & 6 & 18.8 \\
Exploit or security issue with user risk & 4 & 12.5 \\
Censorship or forced inclusion failures & 3 & 9.3 \\
\midrule
\textbf{Total} & \textbf{32} & \textbf{100.0} \\
\bottomrule
\end{tabular}
  }
\end{table}

\section{Discussion}
\label{sec:discussion}

Our analysis in Section~\ref{sec:ethical_analysis} combines two inputs. Table~\ref{tab:l2b_risks_analyzed} captures the prevalence of governance driven risk conditions across 129 projects, while Table~\ref{tab:historic_incidents} records incidents from public reports between 2022 and 2025. There is no one to one mapping between these datasets, so we treat them as complementary rather than causal. Some hazards, such as instant upgrades without exit windows or optional state validation at L1, are latent and may be underrepresented in incident logs. Reporting may be incomplete, labels are coarse, and architectures evolve quickly, so these results should be treated as snapshot.

\textbf{Key insights.}The risks identified in Section \ref{sec:ethical_analysis} are primarily governance risks that arise from who controls core components of L2 stacks, for example sequencers, state root proposers, bridges, and upgrade keys. To appreciate the magnitude of these risks we may consider that the Total Value Secured (TVS), aggregate value of cryptoassets deposited in. Only within 'Others' category, which including least risk-safe systems exceeds 40 billion dollars at the moment of the writing (July 2025).\footnote{\url{https://l2beat.com/scaling/tvs}}

Although these risks can be mitigated with technical controls, their root cause lies in how decision authority and accountability are allocated among operators and governors. Different architectures introduce different trust assumptions, yet the categories of risk apply to both optimistic and ZK rollups. This framing aligns with the external validation in Section \ref{sec:external_validation}, where incident data cluster around operator controlled liveness and finalization paths, and with the prevalence of governance driven hazards in Table \ref{tab:l2b_risks_analyzed}. 


From this analysis we can identify three general directions of mitigation strategies along the lines of: (1) developing standardized taxonomies for classifying trust assumptions across diverse architectures; (2) establishing independent audits and platforms for transparent information dissemination; and (3)exploring alternative technical architectures.

\textbf{Specific recommendations.} Layer 2 rollups, both optimistic and ZK types, have become essential for scaling, but many still operate with centralized training wheels that create security and ethical risks.  Section \ref{sec:external_validation} shows that realized incidents concentrate in operator controlled liveness and finalization paths, while Table \ref{tab:l2b_risks_analyzed} records the high prevalence of exit window absence and whitelisted publication roles. Ensuring user autonomy and safety therefore requires deliberately reducing centralized powers over time. 

Different architectural and governance improvements were proposed before to strengthen rollup security. Recent proposals for based rollups, which assign L2 sequencing rights to Ethereum L1 block proposers~\cite{gorzny2026early}, and for shared sequencing networks~\cite{han2025_a_layer}, which provide a common ordering layer to multiple rollups. Both approaches aim to reduce dependence on a single rollup operator for liveness and ordering, and to alleviate some of the Field 2 and Field 7 configurations identified in Section 4. They also add infrastructure and governance dependencies, so their ethical risks should be assessed with the same eRA lens as current architectures.

We focus on concrete but architecture-agnostic technical and governance measures, and we note where emphasis differs for optimistic versus ZK designs.
\begin{enumerate}
\item Open and permissionless verification of state, fraud proofs and validity proofs.

\textit{Optimistic rollups}. Security hinges on open fraud proof submission during the dispute window. Whitelists create a single point of failure, since collusion or downtime can let fraud go unchallenged. The goal is that anyone running a full node can post a fraud proof, with documented bonds, timeouts, and interfaces. This aligns with the Section 5 power analysis, which warns against small groups holding decision power over correctness while users bear the downside. 

\textit{ZK rollups}. The analogous risk is centralized proving. If one operator controls proof generation, liveness and sometimes safety depend on that operator. Decentralize by allowing multiple independent provers, supporting more than one implementation, and making submission permissionless wherever feasible. Cross verification across implementations reduces correlated failure. 

\item Escape hatches. Escape hatches give users a route to safety even when off chain components fail, which respects user ownership and reduces custodian like risk. They must be non disableable and specified in public parameters,  see~\cite{figueira2025practicalrollupescapehatch} for practical design. Section 5.3 shows sequencer liveness and inclusion incidents are the largest share of incidents, so usable forced inclusion paths and withdrawals anchored in L1 are critical. In optimistic rollups, escape mechanisms interact with the fraud proof pipeline. In ZK rollups, as long as data for the last state are available, any competent prover can step in, which again argues for open participation. 

\item Data availability, a cornerstone requirement. Users must be able to reconstruct L2 state. If data are posted on Ethereum, for example calldata or blobs, validating bridges can rely on on chain commitments. If an external DA layer is used, the settlement bridge effectively relies on a DA oracle that attests that full data for a commitment were made public. Minimum requirements include credible attestations, retention guarantees, public sampling, and an on chain publication fallback. These controls protect optimistic challenges and ZK withdrawals alike. Section \ref{sec:ethical_analysis} notes that optional validation and external DA are prevalent and ethically high severity, even if incidents are sparse. 

\item Multiple independent provers, or verifiers, plus formal verification. A single proving stack, whether a fraud proof VM or a ZK circuit, creates correlated risk~\cite{figueira2025practicalrollupescapehatch}. Two or more distinct mechanisms running in parallel, or at least two independent implementations that cross check, materially raise assurance. Formal methods for circuits and critical contracts further reduce bug classes that would silently propagate through finality pipelines. This complements item 1 by broadening participation and reducing reliance on any single actor~\cite{han2025_a_layer}. 

\item Strictly scoped emergency powers, security councils. Security councils can be useful as circuit breakers, but they concentrate power. Scope them narrowly to verifiable on chain emergencies, for example contradictory validity proofs or objectively detectable faults, and limit actions to the minimum needed to restore protocol invariants. Triggers and procedures should be in contracts, public, and auditable. This responds to Section 5 findings on Field 2, beneficiaries and decision makers without commensurate exposure, and reduces discretionary risk imposition on users. 

\item Time locked governance with ample exit windows. Uncontrolled upgrades are a major risk. Adopt a time lock that provides at least 30 days for users to exit under the old rules for any non emergency change that affects verification, withdrawals, or DA. This covers multiple challenge windows in optimistic designs and allows public review of verifier or circuit changes in ZK designs. An exception is a narrowly scoped fix for adjudicable on chain bugs, which may proceed quickly. Table 1 shows the prevalence of instant upgrades without exit windows is high, which justifies prioritizing this control. 

\item Native L1 verification, a longer term path. Enshrining rollups at the protocol level, sometimes described as ''native'' or ''enshrined'' rollups,  reduce off chain trust and narrows the L1 to L2 gap~\cite{gorzny2026early}. This is a longer term direction, and a full assessment of such designs lies beyond the scope of this paper. The near term protections above remain essential to safeguard users now.

\end{enumerate}

\textbf{Implications for prioritization}. Using eRA, we propose a conservative prioritization that does not conflate prevalence with realized frequency and that accounts for normative severity:
\begin{itemize}
	\item Immediate operational focus guided by incident concentration, strengthen sequencer liveness protections and inclusion paths, with verifiable, documented, and tested parameters, diversify or open proposer and proof submission with timeouts that open participation on lapse, make emergency procedures and fallbacks public and testable. These address frequent Field 7 exposures observed in Table 2.
	\item Structural governance focus guided by high prevalence and severity, eliminate instant upgrades by adopting time locked governance with pre-announced exit windows and escape hatches, make L1 state validation mandatory with published and versioned fault proof or circuit specifications, reduce reliance on external data availability or, where used, implement committees with strong guarantees, public sampling, and an L1 backed exit path. These reduce Field 2 discretion that currently allows risk imposition without corresponding exposure or accountability.
\end{itemize}


\section{Conclusion}

Our analysis combines descriptive prevalence with incident records. It is a snapshot, labels across datasets do not map one to one, reporting is incomplete, and architectures evolve quickly. In L2 rollups, user risk concentrates in operator controlled liveness and finalization paths. Across 129 projects, 86.0 percent allow instant upgrades without exit windows, 50.4 percent are vulnerable to proposer failure that can freeze withdrawals, and 13.2 percent lack a credible forced inclusion path. In 32 publicly documented incidents from June 2022 to August 2025, 59.4 percent involve sequencer disruptions and 18.8 percent involve withdrawal or bridge failures. Together these signals show a pattern where users face the sharpest downside when core operators fail or exercise discretion.

Ethically, this pattern reflects two ethically problematic role configurations highlighted by eRA. One combination appears where decision makers and beneficiaries do not share commensurate exposure. Second appears where users are exposed to risk without decision making capacity. These configurations explain the observed imbalance better than any single technical hazard and they apply across optimistic and ZK designs.

\begin{acks}
We would like to thank Jan Gorzny for his valuable comments on this paper. This work was partially supported by the French ANR project ByBLoS (ANR-20-CE25-0002-01), and by the PriCLeSS project granted by the Labex CominLabs excellence laboratory of the French ANR (ANR-10-LABX-07-01).
\end{acks}

\bibliographystyle{ACM-Reference-Format}
\bibliography{references}

\appendix
\newpage
\section{Historic data L2 incidents}
\label{app:historic_data}
\suppressfloats[t]

\begin{table*}[hbp]
\label{tab:l2_incident_dataset_wide_nodesc}
\centering
\setlength{\tabcolsep}{2pt}
\scriptsize

\begingroup
\setlength{\parskip}{0pt}             
\renewcommand{\arraystretch}{0.96}     
\setlength{\aboverulesep}{0.55ex}      
\setlength{\belowrulesep}{0.55ex}      

\begin{tabular}{@{}p{0.2450\textwidth}p{0.2450\textwidth}p{0.2450\textwidth}p{0.2450\textwidth}@{}}
\toprule
 name of (project) & date & link & incident type \\
\midrule
 Arbitrum One & 2022-06-29 & \url{https://twitter.com/arbitrum/status/1542159109511847937} & Sequencer performance degradation \\
 zkSync Era & 2023-04-01 & \url{https://x.com/zksync/status/1642277357368090626} & Sequencer outage (downtime) \\
 zkSync Era & 2023-04-06 & \url{https://cointelegraph.com/news/zksync-era-denies-921-eth-stuck-forever-in-smart-contract} & Withdrawal failure / Bridge issue \\
 Optimism & 2023-04-26 & \url{https://x.com/optimismstatus/status/1651738268744923137} & Sequencer performance degradation \\
 Arbitrum One & 2023-06-07 & \url{https://x.com/ArbitrumDevs/status/1666549905781825539} & Sequencer halt (batch poster failure) \\
 Shibarium & 2023-08-16 & \url{https://decrypt.co/152773/1-7m-ethereum-stuck-shib-layer-2-network-shibarium} & Bridge halt \& L2 downtime \\
 Base & 2023-09-05 & \url{https://x.com/BuildOnBase/status/1699192035553812658} & Sequencer outage (downtime) \\
 Base & 2023-09-05 & \url{https://status.base.org/incidents/n3q0q4z24b7h} & Sequencer outage (downtime) \\
 Starknet & 2023-11-15 & \url{https://community.starknet.io/t/starknet-downtime-post-mortem-november-15-2023/108339} & Sequencer outage (throughput stall) \\
 Arbitrum One & 2023-12-15 & \url{https://status.arbitrum.io/clq6te1l142387b8n5bmllk9es} & Sequencer outage (downtime) \\
 zkSync Era & 2023-12-25 & \url{https://x.com/zkSyncDevs/status/1739237721963651143} & Sequencer outage (downtime) \\
 Optimism & 2024-02-15 & \url{https://status.optimism.io/clsmrawc31448hmlj2ja14ui3} & Sequencer outage (downtime) \\
 Polygon zkEVM & 2024-03-23 & \url{https://forum.polygon.technology/t/polygon-zkevm-recent-network-outage-report/13702} & Sequencer outage (downtime) \\
 Blast & 2024-03-26 & \url{https://x.com/miszke_eth/status/1772946372309737970} & Censorship or forced inclusion failure \\
 Starknet & 2024-04-04 & \url{https://x.com/Starknet/status/1776216981550563788} & Sequencer outage (consensus bug) \\
 Proof of Play Apex & 2024-05-13 & \url{https://x.com/conduitxyz/status/1790065376975552549} & Sequencer halt \\
 Degen Chain & 2024-05-13 & \url{https://x.com/degentokenbase/status/1789944238731297188} & Sequencer halt \\
 Linea & 2024-06-02 & \url{https://thedefiant.io/news/defi/linea-halts-network-after-velocore-exploit} & Sequencer halt (emergency response) \\
 Scroll & 2024-07-05 & \url{https://status.scroll.io/incidents/44k6s4qg6kcs} & Pending transactions reverted (censored \\
 Mode Network & 2024-08-01 & \url{https://github.com/etherfi-protocol/postmortems} & Withdrawal failure / Bridge issue \\
 Optimism & 2024-08-16 & \url{https://x.com/Optimism/status/1824560759747256596} & Withdrawal failure / Bridge issue \\
 Arbitrum One & 2024-09-25 & \url{https://github.com/OffchainLabs/nitro/releases/tag/v3.2.0} & Exploit or security issue with user risk \\
 Soneium & 2025-01-14 & \url{https://x.com/donnoh_eth/status/1879210463952818472} & Censorship or forced inclusion failure \\
 ZKsync Era & 2025-04-13 & \url{https://zksync.mirror.xyz/W5vPDZqEqf2NuwQ5x7SyFnIxqqpE1szAFD69iaaBFnI} & Exploit or security issue with user risk \\
 RARI Chain & 2025-05-05 & \url{https://app.blocksec.com/explorer/tx/arbitrum/0x4eacd17837407047b65635abdfb9d2693b58efa4040f33baca7b9d27271b0a2c} & Sequencer halt \\
 Abstract & 2025-05-14 & \url{https://dashboard.tenderly.co/tx/0xcaefda7f4c6e29f90b34a0b68817feeb9fac3da2cb66538ea15fbeed434a7201/state-diff} & Sequencer halt (liveness failure) \\
 Scroll & 2025-05-26 & \url{https://forum.scroll.io/t/security-council-report-scroll-mainnet-emergency-upgrade-on-2025-05-26/810} & Exploit or security issue with user risk \\
 Phala & 2025-06-04 & \url{https://x.com/SuccinctLabs/status/1929773028034204121} & Exploit or security issue with user risk \\
 BOB & 2025-07-24 & \url{https://app.blocksec.com/explorer/tx/eth/0xa065f636adfc7cdf08007ee81303028fa4daf291279a75a5ae1d3a975acce806} & Withdrawal failure / Bridge issue \\
 ZKsync Era & 2025-07-30 & \url{https://x.com/zksync/status/1951434107575214429} & Sequencer halt (liveness failure) \\
 Base & 2025-08-05 & \url{https://www.coindesk.com/tech/2025/08/06/base-says-sequencer-failure-caused-block-production-halt-of-33-minutes} & Sequencer outage (downtime) \\
 Scroll & 2025-08-09 & \url{https://etherscan.io/tx/0x3367e24b6cb138cea321f4556259660f24aba1b79ccce8f798ed135e28905f17} & Withdrawal delays \\
\bottomrule
\end{tabular}
\endgroup
\end{table*}

\section{Rollup incident types glossary}
\label{app:incident_glossary}
\suppressfloats[t]   

\begin{table}[hbp]
\label{tab:l2_incident_classes_spaced}
\centering
\setlength{\tabcolsep}{2pt}
\renewcommand{\arraystretch}{1.18}
\scriptsize
\begin{tabular}{@{}>{\raggedright\arraybackslash}p{0.20\linewidth}>{\raggedright\arraybackslash}p{0.47\linewidth}>{\raggedright\arraybackslash}p{0.31\linewidth}@{}}
\toprule
 Incident & Summary & Variability \\
\midrule
 Withdrawal failure & An L2 to L1 exit cannot complete because the exit pipeline is blocked, for example paused bridge, missing L1 acceptance, untrusted outputs, or no escape hatch. & Optimistic failures often follow stuck proposals or disabled fraud proofs, ZK failures often follow verifier misconfiguration or missing validity proofs, both require a live bridge and an L1 recognized commitment. \\
\addlinespace[3pt]
 Sequencer outage & The ordering service is unreachable or non functional, so no new transactions enter the canonical order and the head stalls. Causes include software or infrastructure failures, key loss, or posting dependencies. & Both depend on a sequencing endpoint, some stacks offer an L1 queue for forced inclusion after a timeout, effectiveness depends on parameters and operational readiness. \\
\addlinespace[3pt]
 Sequencer performance degradation & Throughput or responsiveness decays without a hard outage, for example high admission latency, intermittent failures, or batch bottlenecks, inflating mempools and distorting fees and app guarantees. & Shared ordering and L1 posting cause the symptom in both, downstream dispute games versus proving change how much lag users observe. \\
\addlinespace[3pt]
 Sequencer halt & Block production or L1 batch posting stops due to batch poster failure or an operator emergency stop, interrupting finality dependent processes including bridging. & L1 anchoring gates withdrawals and cross domain messages in both, transactions may still be accepted, but exits stall until posting or proving resumes. \\
\addlinespace[3pt]
 Bridge halt & The canonical L1 bridge is paused or disabled, blocking deposits and withdrawals regardless of sequencer or prover status, typically via a privileged governance pause. & Identical effect across families, triggers differ, for example dispute pipeline issues versus verifier or circuit issues. \\
\addlinespace[3pt]
 L2 downtime or severe degradation & Network wide loss of service or materially worse performance for most users, often cascading from other classes, with broad impact on admission and progress. & Both require a working ordering path and an L1 anchoring path, the failing component varies, user impact is reduced ability to transact and settle. \\
\addlinespace[3pt]
 Exploit or security issue with user risk & Vulnerabilities or faults threaten assets or state correctness, for example bridge flaws, unsafe upgrades, disabled validation, or circuit and verifier defects, often requiring emergency changes or temporary disables. & Optimistic risk clusters in dispute games, proposals, and upgrades, ZK risk clusters in prover capacity, circuit soundness, and verifier configuration, both depend on a correct L1 check under governance. \\
\addlinespace[3pt]
 Withdrawal delays beyond normal bounds & Exit latency exceeds the documented timeline from initiation to L1 claimability, indicating pipeline stalls or operator gating even if the formal path exists. & Optimistic overruns point to proposal or dispute issues beyond the challenge window, ZK overruns point to prover backlogs or verifier changes, both gate exits on an accepted L1 commitment. \\
\addlinespace[3pt]
 Documented censorship or forced inclusion failures & Sequencer censorship occurs and the forced inclusion mechanism fails to deliver timely inclusion, for example non triggering timeouts or broken L1 to L2 handoff, making theoretical bypasses ineffective. & Both use L1 mediated inclusion, ZK also needs a validity proof to finalize the forced message, which can extend the censorship window if proving is centralized. \\
\addlinespace[3pt]
 Bridge pauses due to risk & Operators suspend bridge functionality due to suspected vulnerabilities, contamination, or third party incidents, removing deposit and withdrawal paths until conditions for re enablement are met. & Same L1 control surface across families, upstream triggers differ by architecture, user effect is a temporary loss of asset movement. \\
\bottomrule
\end{tabular}
\end{table}

\end{document}